\documentstyle[preprint, aps]{revtex}

\begin{document}
\draft
\preprint{UM-ChE-98/605}
\title{Universality of the excess number of clusters and the crossing probability function in three-dimensional
percolation}
\author{Christian D. Lorenz$^{\#}$ and Robert M. Ziff$^{\dag}$}
\address{University of Michigan, Ann Arbor, MI 48109-2136}

\date{\today}
\maketitle
\begin{abstract}

Extensive Monte-Carlo simulations were performed to evaluate the excess number of 
clusters and the crossing probability function for three-dimensional percolation  on the simple
cubic (s.c.), face-centered cubic (f.c.c.), and body-centered cubic (b.c.c.)
lattices. 
Systems $L \times L \times L^{'}$ with $L^{'} >> L$ were studied for both 
bond (s.c., f.c.c., b.c.c.) and site (f.c.c.) percolation.
The excess number of clusters
$\tilde {b}$ per unit length was
confirmed to be a universal quantity with
a  value $\tilde {b} \approx 0.412$.
Likewise, the critical crossing probability in the $L^{'}$ direction,
with periodic boundary conditions
in the $L \times L$ plane,  was found to follow a universal exponential decay 
as a function of $r = L^{'}/L$ for large $r$.  Simulations were also carried out to find new precise values of the critical
thresholds for site percolation on the f.c.c.\
and b.c.c.\ lattices, yielding $p_c(\rm {f.c.c.})= 0.199\,236\,5 \pm
0.000\,001\,0$,  $p_c(\rm {b.c.c.})=
0.245\,961\,5\pm 0.000\,001\,0$.  We also report the value $p_c(\rm{s.c.}) = 0.311\,608\,0 \pm 0.000\,000\,4$ for site
percolation.
\end{abstract}

\narrowtext
\section{Introduction}
\label{sec1}

The standard percolation model \cite{SA} involves the random
occupation of sites or bonds of a regular lattice.
At a critical occupation probability $p_c$,
the mean size of clusters of occupied
sites becomes infinite, while the number of clusters $n(p)$
per site or per unit volume 
remains finite with $n_c =
n(p_c)$.  

The value of $n_c$ depends on the microscopic characteristics of each
system, and because of this it is a non-universal quantity.  For two-dimensional (2d) systems, precise numerical values of
$n_c$ for
bond and site percolation on the square and triangular
lattices were found by Ziff, Finch,
and Adamchik \cite{ZFA}, whose results for bond percolation confirmed the theoretical
predictions of Temperley and
Lieb \cite{TL} and Baxter, Temperley, and Ashley \cite{BTA}.
 In 3d,
there are no theoretical predictions for $n_c$, and its values for
different systems apparently have not been reported
in literature.  

In Ref. \cite{ZFA}, it was also found that the {\it excess} number of clusters
$b \equiv \lim_{L\to
\infty} LL^{'}(n(L,L^{'}) - n_c)$, with $r = L^{'}/L =$ fixed,
where $n(L,L^{'})$ is the number of clusters per unit area in
a finite system of size $L \times L^{'}$
with periodic boundary conditions, is a universal quantity that
depends only upon aspect ratio $r$. (Note that  in \cite{ZFA}, the authors defined $n$ as 
clusters per site rather than per unit area, but the result for $b$ is the same.)
 This universality is consistent with the arguments
of Privman and Fisher \cite{FP},
and has also been discussed by Aharony and Stauffer \cite{SA2} and  by M\"{u}ller \cite{Muller} for the
Ising model.  Kleban and Ziff
\cite{KZ} introduced
an excess number per unit length $\tilde{b} \equiv \lim_{r \to \infty} b(r)/r 
= \lim_{L\to \infty}
L^{2}(n(L,L^{'})-n_c)$ for long
cylindrical systems
$L^{'}>> L$, and derived exact results for both $b(r)$ and $\tilde{b}$
in 2d systems.
Again, however, no theoretical predictions for $b$ in 3d exist. 

In this paper, we determine $n_c$ and $\tilde{b}$ for various 3d rectangular solid systems of dimensions $L
\times L \times L^{'}$ with
$L^{'} >> L$.  We consider bond percolation on the simple cubic (s.c.),
body-centered cubic (b.c.c.),
and face-centered cubic (f.c.c.) lattices, and site percolation on the
f.c.c.\ lattice.

A prerequisite to
finding the value of
$n_c$ for each of these systems is knowing the 
critical occupational probability
$p_c$ to high accuracy.   Previously, accurate values
were found for bond percolation on all three lattices and
site percolation on the s.c.\ lattice only, as summarized in Table \ref{table1}.
  To round out these values, we carried out simulations to determine $p_c$ for site percolation on the 
b.c.c.\ and f.c.c.\ lattices to high accuracy --- although we used only the latter in the study of the
excess cluster number, since the universality was clearly confirmed
with the four systems that we studied.  In another work \cite{LZ2}, we have studied site percolation on the s.\ c.\
lattice, and report this result in Table \ref{table1} also.

The simulations for finding $n_c$ were also used to study
the critical crossing probability for the three-dimensional systems.
The crossing probability function $\pi(\Gamma)$ of a system of shape $\Gamma$ gives the probability that at least one 
cluster connects two disjoint pieces of the boundary $\partial \Gamma$,
and has been of much interest lately 
following the realization that it is a fundamental, universal property of percolation,
independent of the underlying
lattice type, and subject to conformal invariance \cite{Langlands,Cardy,Ziff92,Aizenman}.
  In 2d, Cardy \cite{Cardy} derived an explicit expression for the 
vertical crossing probability $\pi_v$ of rectangular systems $L \times L^{'}$, with open 
boundaries in the horizontal direction, and Watts \cite{Watts} derived an expression for the probability
of vertical but not horizontal crossing for this system.  The $\pi_v$ for 2d systems with 
periodic (and other) boundary conditions was studied by Hovi and Aharony \cite{HA}. 
A number of systems were also studied by various groups including Hu {\it et al.}\ \cite{Hu}, Hsu {\it et al.}\
\cite{HSU}, Gropengiesser
\cite{gropp}, and Vicsek and Kert\'{e}sz \cite{VK} .  In 3d, work has been restricted to simple cubical boundaries $L
\times L
\times L$, with crossing studied between two opposite planes and various boundary conditions on the sides
\cite{Hu,Adler,ST}.

Here we find $\pi_v$ for the $L \times L \times L^{'}$ systems for all $L^{'}$ by measuring the distribution of the maximum
height of clusters connected to the base of the rectangular system.
(A similar method was used in \cite{Zel} for 2d systems.)
We consider crossing in the $L^{'}$ direction for systems with periodic
boundary conditions in the $L \times L$ plane, and show that $\pi_v$ is
a universal function of $r = L^{'}/L$ for large $L$.

In the following three sections we report on the determination of the new values of
$p_c$, the determination of 
$n_c$ and $\tilde{b}$, and 
the determination of $\pi_v(r)$.  The results are summarized and discussed further in the conclusions section.

\section{Percolation thresholds}
\label{sec2}

Precise values for the thresholds for bond percolation on 
all three lattices, and for site percolation on the s.c.\ lattice,
 have been found elsewhere.
Here we also determine accurate values for site percolation 
on the f.c.c.\ and b.c.c.\ lattices.
A summary of our results and other recent results is given in Table \ref{table1}.  

The procedure we used to find $p_c$ was similar to
that we used for bond percolation in \cite{LZ}.  We grew individual clusters by a
Leath-type algorithm and identified the critical point using an epidemic scaling analysis.
A virtual lattice of $2048^3$ sites was simulated, using the block-data
method first described in  \cite{ZCS}.
There were only two minor changes made to the simulation of \cite{LZ} so that it could
be used to study site
percolation.  First, as the clusters were grown, the sites were either
occupied with a probability,
$p$, or left vacant with a probability, $1-p$.  If a site was determined to be
vacant, then (unlike in bond percolation) it was
never revisited as a potential growing site.
The other difference is the
cut-off for the growth of these clusters was set to
$2^{19}$
(524,288) wetted sites, as opposed to $2^{20}$ (1,048,576) and $2^{21}$
(2,097,152) in ref.
\cite{LZ}.

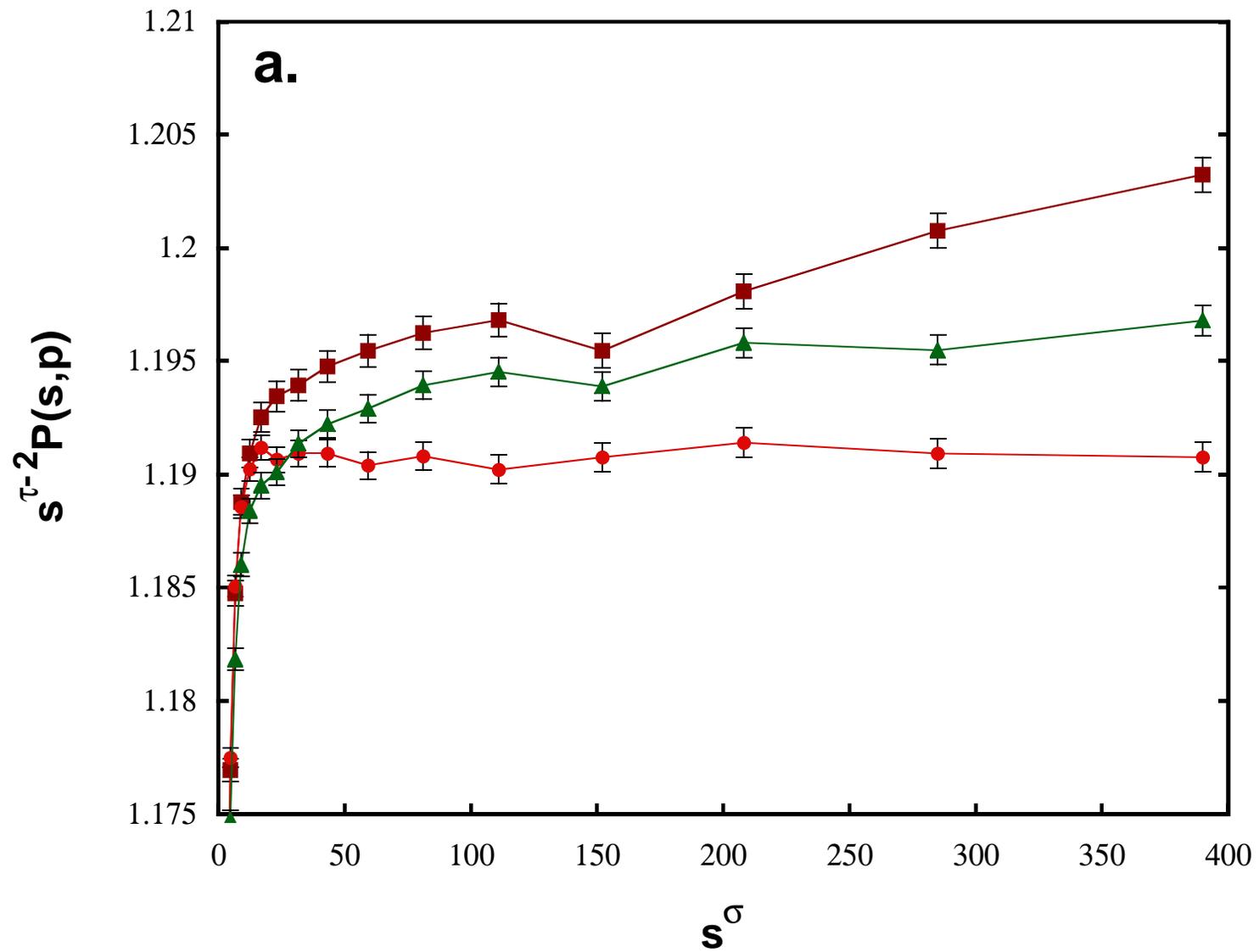
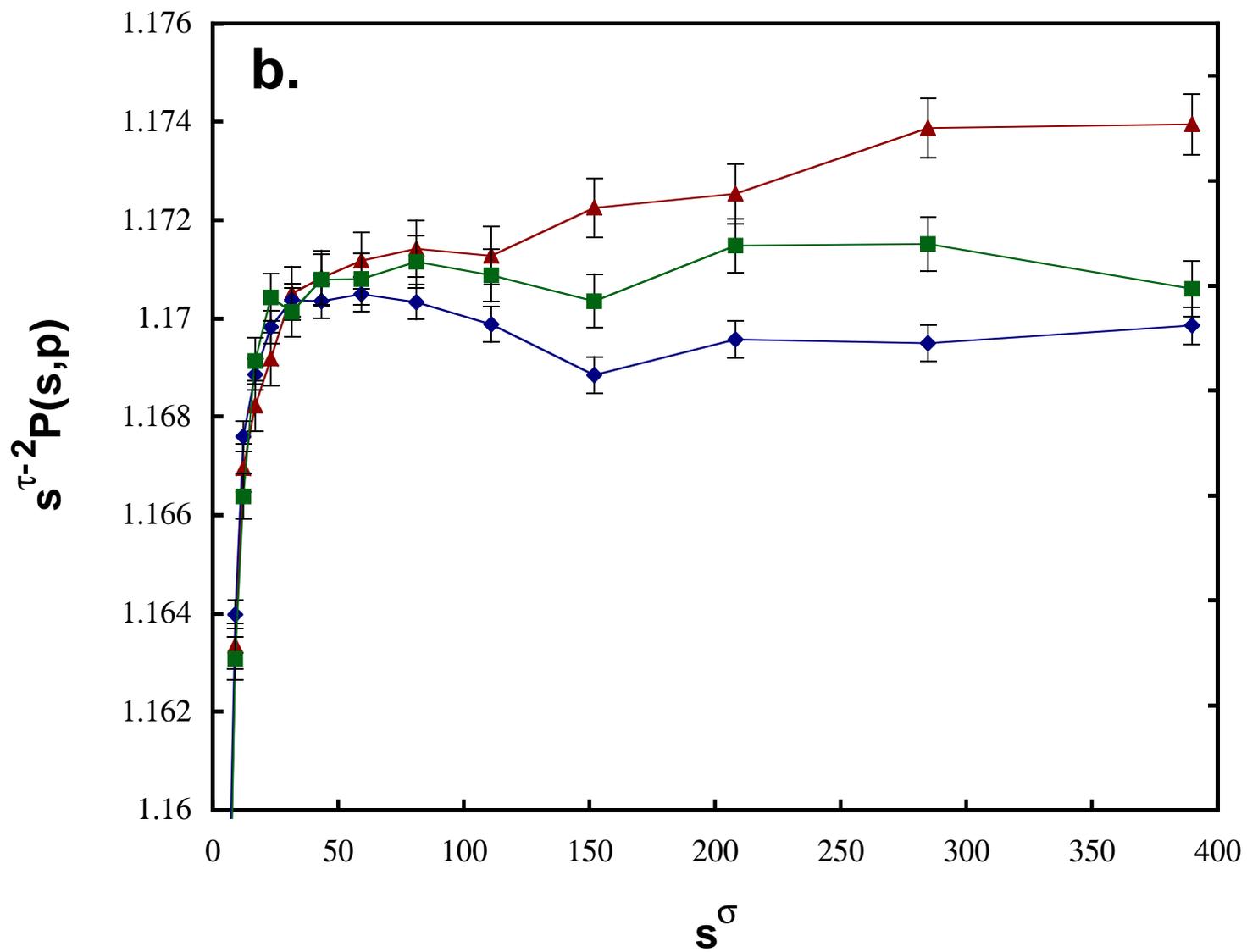
\begin{figure}
\caption{ Plot of $s^{\tau-2}P(s,p)$ vs.\ $s^\sigma$ for
(a) f.c.c.\ and (b) b.c.c.\ lattices using $\tau = 2.189$ and $\sigma = 0.445$.  Each curve represents a different value of
$p$, which are (from top to bottom) (a) $0.199\,237\,5$, $ 0.199\,236\,5$, and $
0.199\,235\,5$, and (b)
$0.245\,962\,5$,
$0.245\,961\,5$, and $0.245\,960\,5$.
\label{figurepc}}
\end{figure}

The simulation yielded the fraction 
of clusters $P(s,p)$ that grew to a size greater than or equal to $s$ sites.
When $p$ is near $p_c$, one expects $P(s,p)$ to behave as
\begin{equation}
P(s,p) \sim As^{2- \tau} f\bigl((p-p_c)s^\sigma \bigr) \approx 
  As^{2- \tau} [ 1+C(p-p_c)s^\sigma+\ldots ]
\label{Taylor}
\end{equation}
where $\tau$ and $\sigma$ are universal exponents \cite{F}.
Here we assumed the values   $\tau = 2.189$ and $\sigma = 0.445$, consistent
with other 3d work \cite{LZ,ZS,Betal}.  As in \cite{LZ}, plots of
$s^{\tau - 2}P(s,p)$ vs.\ $s^\sigma$
for site percolation of the b.c.c.\ and f.c.c.\ lattices were used to find
the value of the percolation
threshold, which corresponds to horizontal behavior for large $L$ on such a plot.
In all, we generated $1.5\times10^{7}$ clusters for the
f.c.c.\ lattice and  $2.2
\times10^{7}$ for the b.c.c.\ lattice for a range of values of $p$ requiring several weeks of workstation computer time. 
The results are plotted in Fig.\ \ref{figurepc} and imply the following values for the critical thresholds:
\begin{eqnarray}
\label{percthresh}
 && p_c(\rm{b.c.c.}) = 0.245 \, 961 \, 5 \pm 0.000 \, 001 \, 0 \nonumber \\
&&p_c(\rm{f.c.c.}) = 0.199 \, 236 \, 5 \pm 0.000 \, 001 \, 0 
\end{eqnarray}
These results were consistent with (and more than 1000 times more precise
than) previous work, as shown in  Table \ref{table1}.

\section{Values of $n_c$ and the finite-size correction $\tilde{b}$}
\label{sec3}
Using the values of the critical thresholds given in Table \ref{table1}, we carried out simulations
to measure the number of clusters for bond percolation on each of the three-dimensional lattices and site
percolation on the f.c.c.\ lattice.  (We did not consider site percolation on the s.c.\ and b.c.c.\ lattice
in this calculation.)
 Clusters were grown successively 
from every unvisited site by a growth algorithm
\cite{L} on a three-dimensional square bar, $L \times L \times L^{'}$
with $L^{'} >> L$.  Periodic 
boundary conditions were assumed in each horizontal plane. (Here, vertical is taken to be the
$L^{'}$ direction).  The
first cluster was started in the upper left-hand corner of the first plane
($z=0$) at the point (0,0,0). 
From this corner, a cluster was grown to the nearest neighboring sites as
defined for each system by the unit vectors in
\cite{LZ}, occupying the connecting bonds or neighboring sites with a probability,
$p_c$, and leaving them unoccupied with a probability,
$1-p_c$.  After the first cluster was grown, a new cluster was seeded from
the first unoccupied site
in the left-most column, and grown until it died.  After all sites of
the first plane were tested,
the growing plane was moved to $z=1$, and so on.
Because the previous planes were completely occupied, their data
could be discarded and the
memory recycled.  Furthermore, the clusters never extended up to a plane
of distance $z=32L$ from the
growing plane.  As a consequence, a system of size
$L
\times L
\times 32L$ could be used to effectively simulate a
$L \times L \times \infty$ system by wrapping around in the third direction.  

We ran simulations to $L^{'}= 2\times
10^{9}$, with
$L$ = 4, 5, 6, 7, 8, 10, and 12 for the s.c.\ lattice, $L$ = 4, 6, 8, 10
and 12 for the f.c.c.\ lattice (both site and bond), and
$L$ = 4, 6, 8, 10, 12, and 16 for the b.c.c.\ lattice.
 In total, we grew about $1.06\times10^{12}$ clusters for the b.c.c.\ lattice, $1.88\times10^{12}$
clusters for the s.c.\ lattice,
$1.44\times10^{12}$ clusters for bond percolation on the f.c.c.\ lattice,
and $3.32\times10^{11}$ for
site percolation on the f.c.c.\ lattice, which required several additional months of computer time.  

In Figure
\ref{figure1}, we display a representative $4 \times 4$ plane of each of the
three
lattices, showing how the lattices were oriented in our simulations and how
the unit dimension was defined. The
darkened circles represent active sites  in the current plane, and the
empty circles represent active
sites that are in the neighboring planes.  The solid
lines are bonds which lie
within the current plane and the dashed lines represent bonds which
connect the displayed plane to the
neighboring planes.  For modeling the s.c.\ lattice, the
plane shown in Figure
\ref{figure1}(a) is repeated for the whole length of the cylinder, while
for the other two lattices, the plane shown in the figure is repeated
on every other plane.   In the
case of the s.c.\ lattice, all
of the available sites in the plane are considered active, for the f.c.c.\ lattice,
only half of the underlying cubic-lattice sites are active,
and for the b.c.c.\ lattice, only a quarter of the cubic-lattice sites are active.
Note that the unit dimension that we define for the b.c.c.\ and f.c.c.\
lattices is neither the unit cell dimension nor the nearest-neighbor
distance, but half of the unit cell dimension.

Now, for a finite system of volume $V$ with periodic boundary
conditions, analogous to what was found in \cite{ZFA} for 2d, we expect 
\begin{equation}
n = n_c + { b \over V} + { c \over V^2} + \ldots
\label{correction}
\end{equation}
where $b$, representing the excess number of clusters in this finite
system, is universal, a function of the shape only.  
Here we studied $L \times L \times L^{'}$ systems, where $L^{'} >> L$,
with the volume given
by $V = L^2 L^{'}$. \ For systems of this shape, we expect the excess
number of clusters per unit length $b /  (L^{'}/L) $ to be a constant $\tilde{b}$, i.e.,
\begin{equation}
b \sim \tilde{b}L^{'}/L
\label{beq}
\end{equation}
for $L^{'} >> L$.  Likewise, we write $c \sim
\tilde{c}(L^{'}/L)^2$.  Then it follows from (\ref{correction}) that 
\begin{equation}
n = n_c + {\tilde{b} \over L^3} + { \tilde{c} \over L^6} + \ldots
\label{correctionL}
\end{equation}
 Both $b$ and
$\tilde{b}$ are functions of the system shape only and are  universal
quantities, but $c$
and
$\tilde{c}$ vary from system to system and are not universal.
Eq.\ (\ref{correctionL}) implies that $n_c$ can be found from a plot of $n$ vs.\ $1/L^3$, as shown in Figure
\ref{figure2} for our data from the b.c.c.\ lattice.  The values of $n_c$, which are shown in units of number of
clusters per unit volume as
defined in Fig.\ \ref{figure1}  for the various lattices, are given in Table
\ref{table2}.  These values can be converted to units of number of clusters
per site by taking into account
that the s.c., f.c.c., and b.c.c.\ lattices have 1, 1/2, and 1/4 sites per
unit volume, respectively.

Equation (\ref{correctionL}) can be rearranged as
\begin{equation}
(n - n_c)L^3 = \tilde{b} + {\tilde{c} \over L^3} + \ldots
\label{correction2}
\end{equation}
Therefore, once $n_c$ is determined, $\tilde{b}$ and $\tilde{c}$ can be found from a plot of $(n
-n_c)L^3$ vs.\ $1/L^3$.  Figure \ref{figure3} shows
this plot for the systems that
we studied.  The resulting values of $\tilde{b}$ and  $\tilde{c}$ for each of the
systems are listed in Table
\ref{table2}.  A universal value of $\tilde{b} = 0.412 \pm 0.002$
is obtained from these results.

\section{Critical crossing probability $\pi_v$}
\label{sec4}

 Our simulations for $n_c$ could also be
used to obtain $\pi_v$ by comparing the distance from the growth plane to the maximum height
plane.  If this distance is greater
than or equal to some fixed value $L^{'}$, then crossing will occur in an
$L \times L \times L^{'}$
system (with periodic boundary conditions in each $L \times L$ plane).
In other words, we could determine $\pi_{v}(L, L,
L^{'})$ for all $L^{'}$ by keeping track
of the distribution of distances between the growth plane and maximum
height planes in our continuous simulations.

In 2d, the probability of crossing a
system of aspect ratio  $r$ = height/width in the vertical direction,
with periodic boundary
conditions in the horizontal direction,
is given by \cite{Aizenman,cardy98}
\begin{equation}
\pi_v(r) \sim e^{-2 \pi r (2-D)} = e^{-{5 \over 24} \pi r} 
\label{ccp}
\end{equation}
for large $r$, where $D=91/48$ is the 2d fractal
dimension.  Eq.\ (\ref{ccp}) follows from a conformal
transformation from an annulus to a rectangle, using that the probability a cluster extends beyond a radial
distance  $R$ scales as $R^{D-d}$.  We have separately verified
that Eq.\ (\ref{ccp}) holds accurately for all $r$ somewhat greater than 1.
 
For 3d systems, while it still is true that the radial
probability scales as $R^{D-d}$, we cannot connect it to $\pi_v$ of the $L \times L \times L^{'}$ system, because we
cannot make a conformal  transformation between the concentric spheres and a rectangular
solid.  However, we still expect an exponential dependence upon $r = L^{'}/L$, because
that term represents the smallest eigenvalue of the 
transfer matrix.  We thus hypothesize
\begin{equation} 
\pi_v \sim Ke^{-mr}.
\label{ccpln}
\end{equation}  
for large $r$.  To check this, we plot $\ln \pi_v$ vs.\
$r$ in Fig.\ \ref{figure4}, which 
contains the results from all four
systems studied, for $L =$ 8, 10, and 12.  To get the best data collapse,
we defined $r = (L^{'} + \ell)/L$, which allows for a lattice finite-size
effect or boundary extrapolation length in the $L^{'}$ direction, in which the effective location of the free boundary is
not uniquely defined \cite{Zel}. (Such an ambiguity in size does not occur in the $L$ directions,
because of the periodic boundary conditions.)  
In fact, the data for all three bond percolation systems collapsed
nicely with $\ell = -1.3$, while the data for site
percolation on the f.c.c.\ lattice required a constant of $\ell = 1.36$ to fall on the same curve.
Figure \ref{figurecollapse} shows the effect of $\ell$ by comparing an enlarged portion of our data from the s.c.\ (bond)
lattice when
$\ell = 0$ and $\ell = -1.3$.  The corresponding values of $m$ and $\ln K$ are $-1.37 \pm 0.01$ and $0.75 \pm 0.05$,
respectively.

\section{Discussion of results}
\label{sec5}

Our values for the critical thresholds of site percolation on the f.c.c.
and b.c.c. lattices are listed in Table \ref{table1}.  Along with the other results
which are 
summarized in that table, the thresholds of all three 3d systems, for
both site and bond percolation, are now known to a very high accuracy.

Table \ref{table2} lists $n_c$, $\tilde b$ and $\tilde c$ for the four systems
studied.  
Our simulations confirm that $\tilde{b}$ is universal in
3d as it is in 2d \cite{ZFA}, 
with a value $\tilde b \approx 0.412$.  In 2d, the corresponding value is $\tilde{b} = 5\sqrt{3} / 24 =
0.360\,844\ldots$
\cite{KZ}.

The average density of clusters per site, $n_c$, varies from system
to system, as expected.
The
values for $n_c$ in Table \ref{table2} show that the simple cubic is the
most dense system, according to the convention we used to define the unit
volume of the system.

Our simulations have also shown that $\pi_v$ is universal as shown in
Figure \ref{figure4}, and possesses an exponential decay (\ref{ccpln})
with $m = 1.37 \pm 0.01$, compared with a value of $5 \pi/24 = 0.654\,498\ldots$
in 2d.  For a  cubical system ($L \times L\times L$ or $r=1$), Eq.\ (\ref{ccpln}) implies a value of $\pi_v = 0.54 \pm
0.04$, while a direct analysis of our data at that point yields the more precise value
$\pi_v = 0.573 \pm 0.005$.  The latter
value is somewhat higher than the result $0.513 \pm 0.005$ recently reported by Acharyya and Stauffer \cite{ST} for a
system with helical boundary conditions in the plane, which are similar to periodic boundary conditions but
with the rows shifted by one.  We believe that in the limit of large $L$ these two boundary conditions should be
equivalent, although this belief is not supported by the discrepancy in the values seen above.

Many additional questions are raised for 3d systems.  What is $b(r^{\prime}, r^{\prime\prime})$ where $r^{\prime}=L^{'}/L$
and $r^{\prime\prime}=L^{''}/L$ for an $L \times L^{'} \times L^{''}$ system (with periodic boundary conditions in all
directions)?  What is the effect of helicity or a twist of the order $L$ in the periodic boundary conditions?  Is
$\tilde{b}$ related to the number of ``percolating'' clusters per unit length (however precisely that may be defined)? 
Finally, can one devise a system that conformally
transforms to concentric spheres, so that the crossing probability across that system will be
given by a formula analogous to (\ref{ccp})?

\acknowledgments

This material is based upon work supported by the US 
National Science Foundation under Grant No.\thinspace DMR-9520700.  Dietrich Stauffer is thanked for useful comments.\\

$^{\#}$Electronic mail: cdl@engin.umich.edu

$^{\dag}$Electronic mail: rziff@engin.umich.edu

\begin{figure}
\caption{ 
Representative $4 \times 4$ planes for the (a) s.c., (b) f.c.c., and (c)
b.c.c.\ lattices.  The
darkened circles represent active sites in the plane and empty circles
represent active sites in the
neighboring planes.  The solid lines represent bonds in the
plane and dashed lines represent
bonds which go to the neighboring planes.
\label{figure1}}
\end{figure}

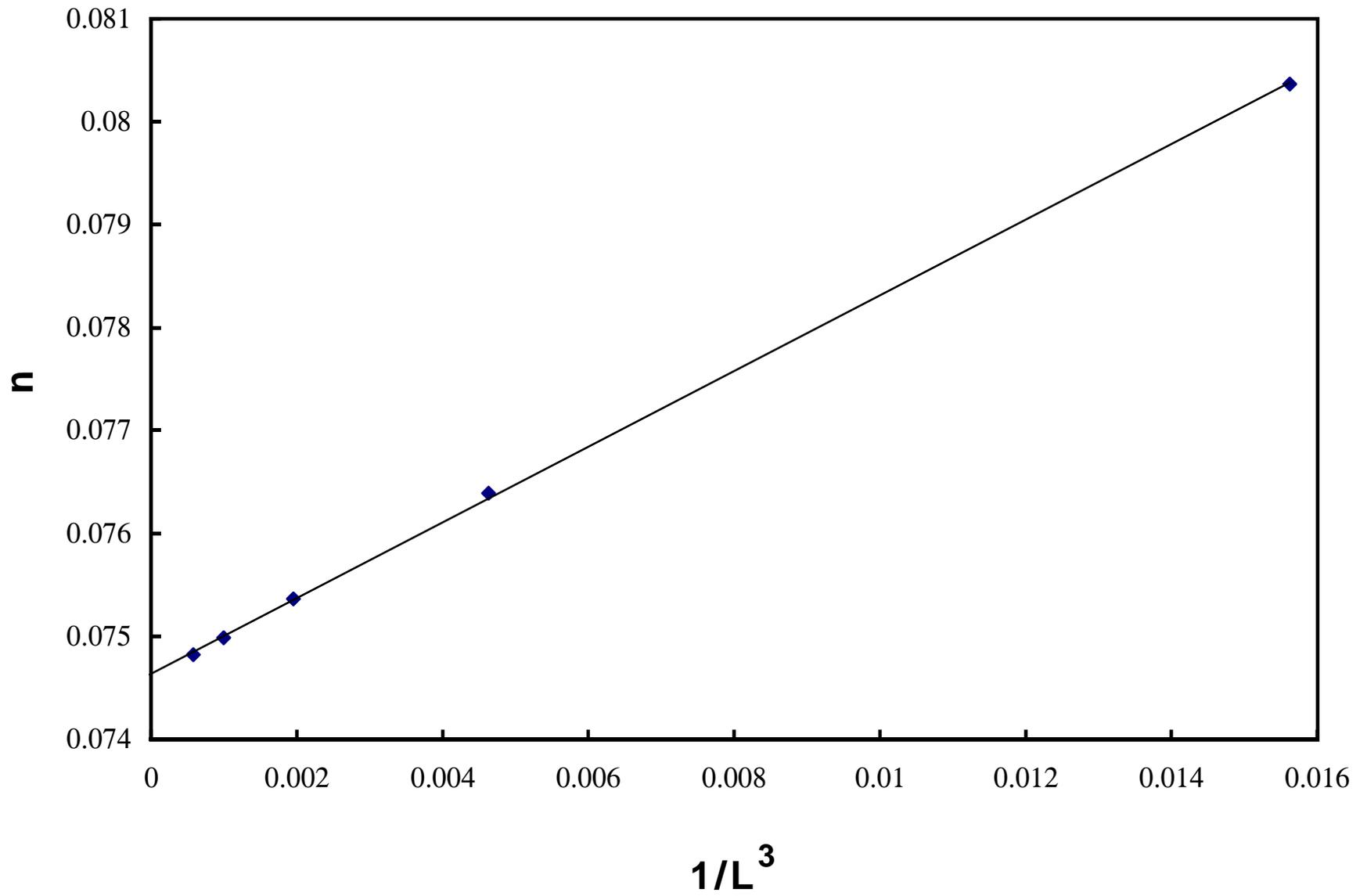
\begin{figure}
\caption{ 
Plot of $n$ vs.\ $1/L^{3}$ for bond percolation on the b.c.c. lattice.  The intercept of this plot
yields $n_c$ and the slope
yields $\tilde{b}$ according to Eq.\ (\ref{correctionL}).
\label{figure2}}
\end{figure}

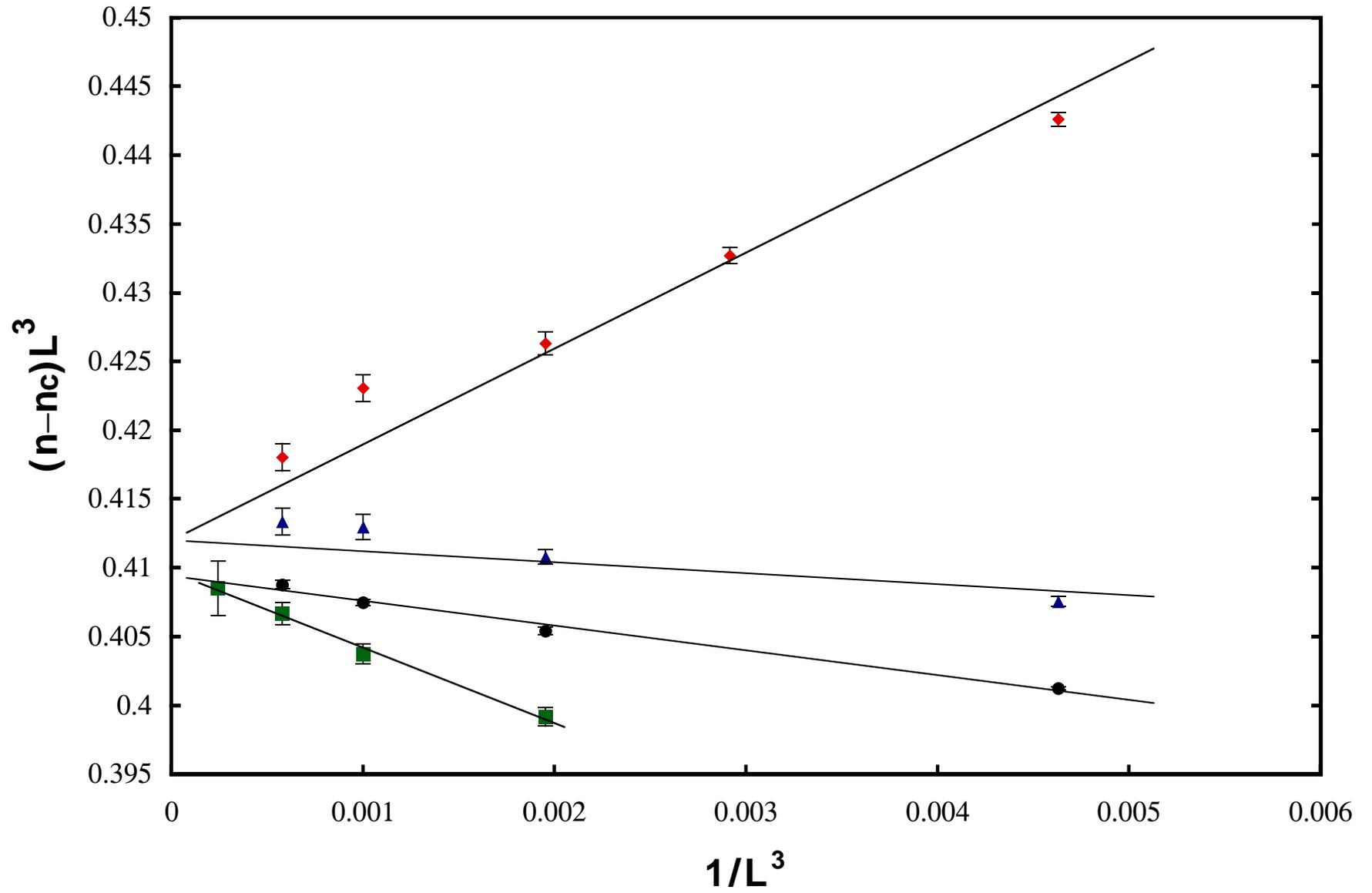
\begin{figure}
\caption{Plot of $(n-n_c)L^{3}$ vs.\ $1/L^{3}$ for the b.c.c.\ (bond), f.c.c.\ (bond), f.c.c.\ (site), and s.c.\ (bond)
(from top to bottom) systems at
$p_c$. In these plots, the intercept represents the value of $\tilde{b}$
and the slope is the second
correction term $\tilde{c}$.  The values of $\tilde{b}$ and $\tilde{c}$
are listed in Table \ref{table2}.
\label{figure3}}
\end{figure}

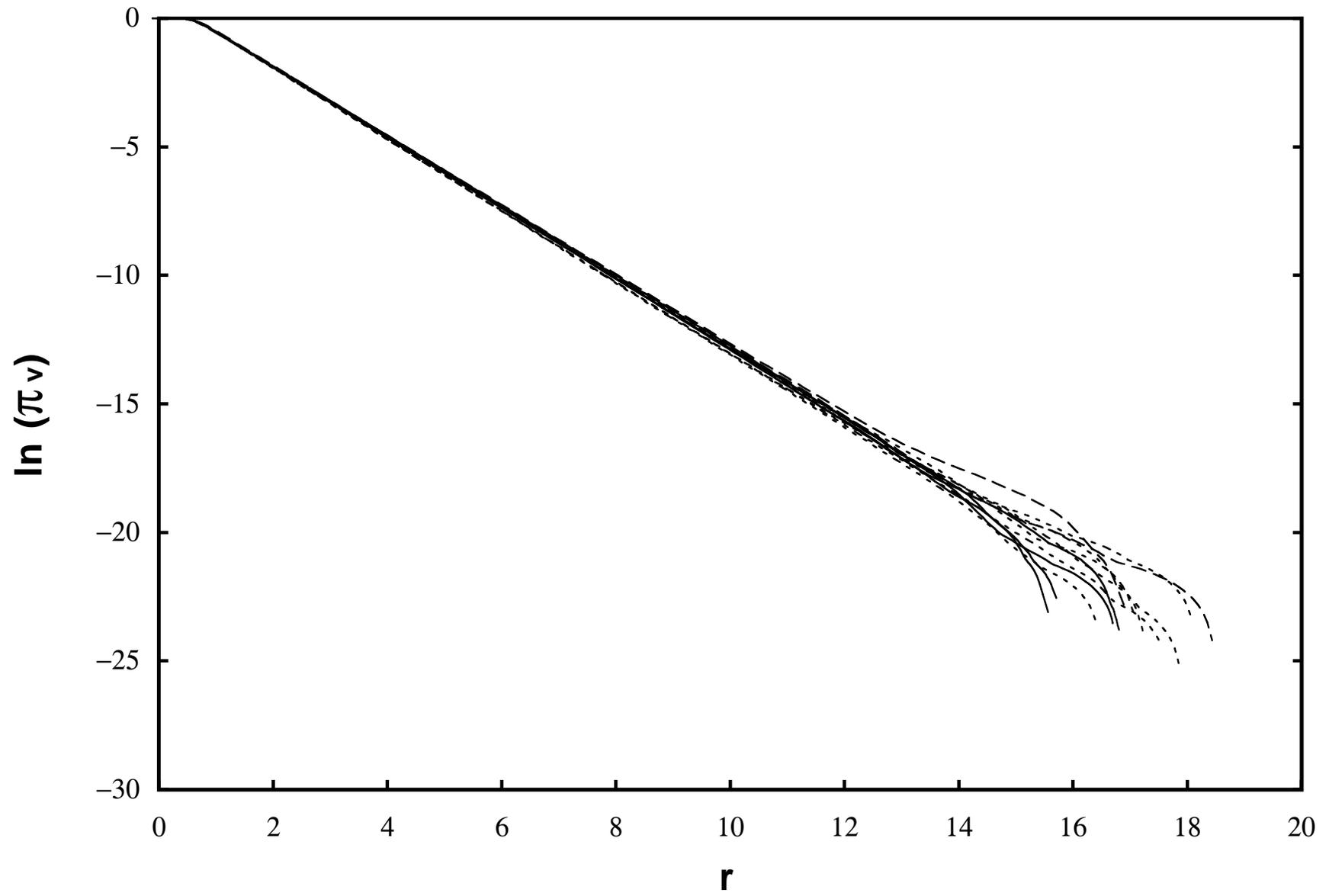
\begin{figure}
\caption{Plot of $\ln \pi_v$ vs.\ $r=(L^{'}+\ell)/L$ for the b.c.c.\ (bond) (dashed lines), f.c.c.\ (bond) (dotted lines),
f.c.c.\ (site) (also dotted lines), and s.c.\ (bond) (solid lines) lattices of size
$L\times L\times
L^{'}$ with $L =$ 8, 10, and 12 at
$p_c$. In these plots, the intercept represents the value of $\ln K$ and the
slope is $m$.  The values of
$\ln K$ and $m$ are $0.75 \pm 0.05$ and $-1.37 \pm 0.01$.
\label{figure4}}
\end{figure}

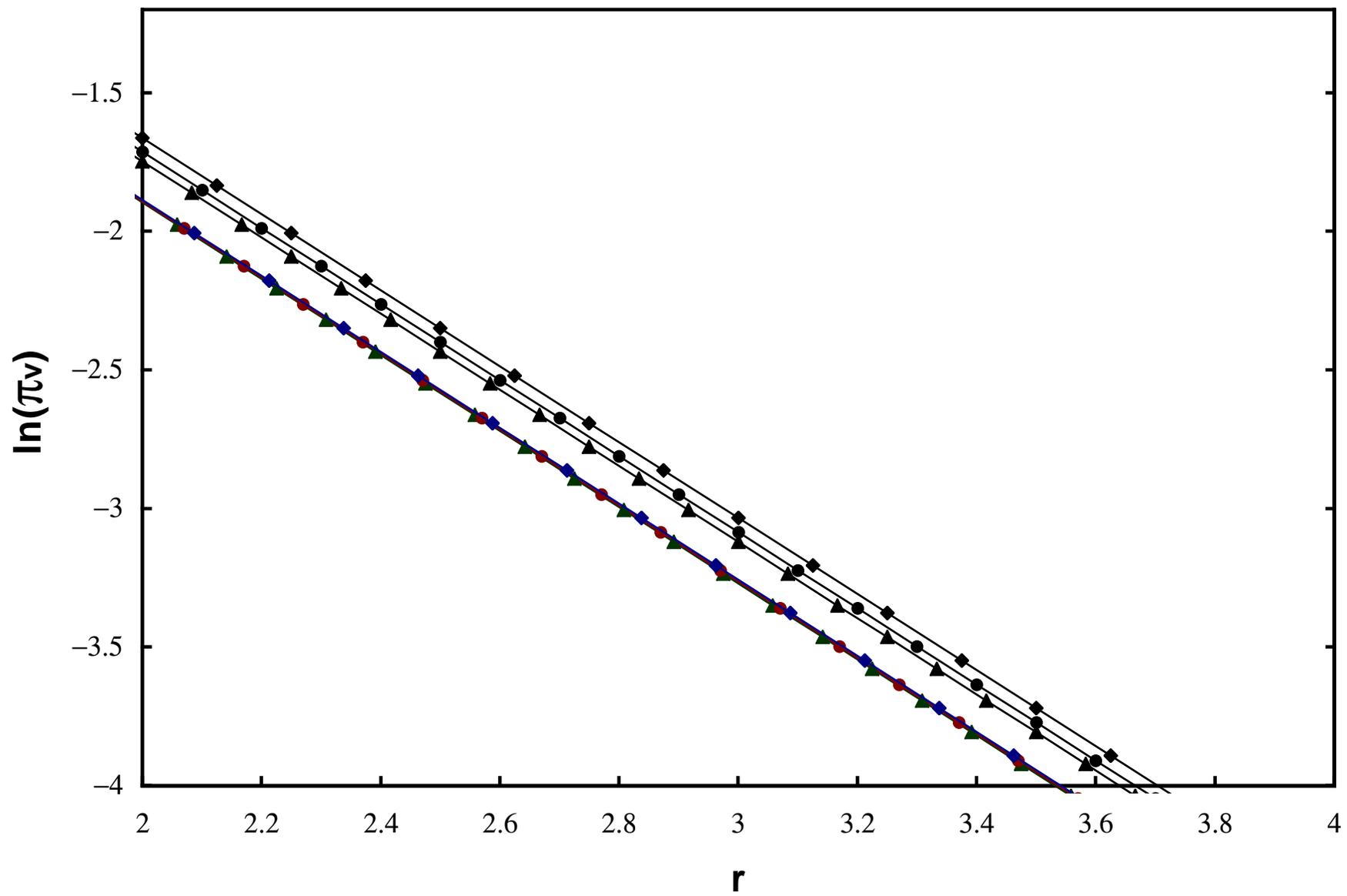
\begin{figure}
\caption{Plot of $\ln \pi_v$ vs.\ $r=(L^{'}+\ell)/L$ for an enlarged portion of the data from the s.c.\ (bond) lattice of
size
$L\times L\times
L^{'}$ with $L =$ 8 (square), 10 (circle), and 12 (triangle) at
$p_c$. The upper three curves show the data plotted with $\ell=0$, and the bottom curve shows the data collapse when
$\ell=-1.3$ is used.
\label{figurecollapse}}
\end{figure}

\vfill\eject

\begin{table}
\caption{Values of $p_c$ for bond and site percolation on the b.c.c., f.c.c.\ and s.c.\ lattices from present (*) and
other recent work.  The numbers in parenthesis represent the errors in the last digit(s).
\label{table1}}
\begin{tabular}{llll}
system&$p_c$&Ref.&Value used here\\
\tableline
\tableline
b.c.c.\ (bond)&$0.180\,3$&\cite{SA}&\\
&$0.180\,2(2)$&\cite{vanderMarck}&\\
&$0.180\,287\,5(10)$&\cite{LZ}&$0.180\,287\,5$\\
b.c.c.\ (site)&$0.246$&\cite{SA}&\\
&$0.245\,8(2)$&\cite{vanderMarck}&\\
&$0.245\,961\,5(10)$&*\\
\tableline
f.c.c.\ (bond)&$0.119$&\cite{SA}&\\
&$0.120\,0(2)$&\cite{vanderMarck}&\\
&$0.120\,163\,5(10)$&\cite{LZ}&$0.120\,163\,5$\\
f.c.c.\ (site)&$0.198$&\cite{SA}&\\
&$0.199\,4(2)$&\cite{vanderMarck}&\\
&$0.199\,236\,5(10)$&*&$0.199\,236\,5$\\
\tableline
s.c.\ (bond)&$0.248\,8$&\cite{SA}&\\
&$0.248\,7(2)$&\cite{vanderMarck}&\\
&$0.248\,8(2)$&\cite{AMHA}&\\
&$0.248\,75(13)$&\cite{Grass}&\\
&$0.248\,814(3)$&\cite{G}&\\
&$0.248\,812(2)$&\cite{ZS}&\\
&$0.248\,812\,6(5)$&\cite{LZ}&$0.248\,812\,6$\\
s.c.\ (site)&$0.311\,6$&\cite{SA}&\\
&$0.311\,4(4)$&\cite{vanderMarck}&\\
&$0.311\,605(10)$&\cite{ZS}&\\
&$0.311\,604(6)$&\cite{G}&\\
&$0.311\,605(5)$&\cite{ST}&\\
&$0.311\,600(5)$&\cite{JS}&\\
&$0.311\,608\,1(13)$&\cite{Betal}&\\
&$0.311\,608\,0(4)$&\cite{LZ2}&\\
\end{tabular}
\end{table}

\begin{table}
\caption{Values of $n_c$ (clusters per unit volume), $\tilde{b}$, and $\tilde{c}$ for the systems studied.
\label{table2}}
\begin{tabular}{llll}
system&$n_c$&$\tilde{b}$&$\tilde{c}$\\
\tableline
\tableline
s.c.\ (bond)&$0.272 \, 931 \, 0(5) $&$0.414(3)$&$6.0(7)$\\
\tableline
f.c.c.\ (bond)&$ 0.153 \, 844 \, 0(5) $&$0.414(3)$&$-1.4(3)$\\
f.c.c.\ (site)&$ 0.013 \, 265 \, 5(5) $&$0.409(3)$&$-1.8(3)$\\
\tableline
b.c.c.\ (bond)&$0.074 \, 586 \, 0(5)$&$0.409(3)$&$-5.5(7)$\\
\end{tabular}
\end{table}
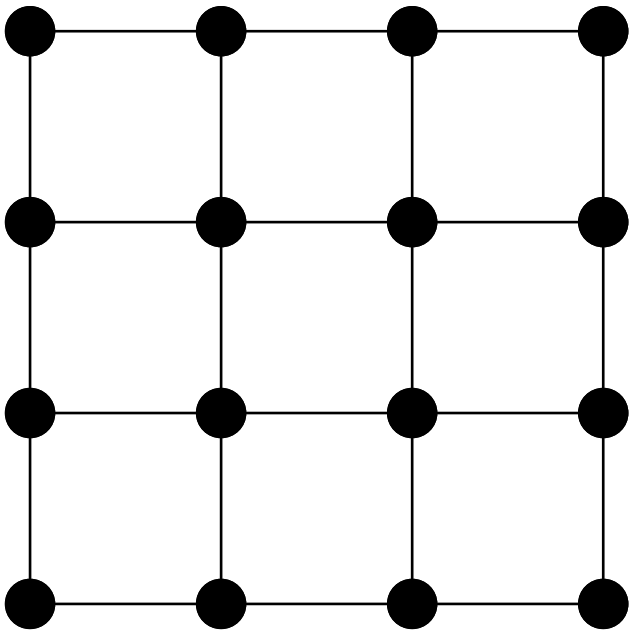
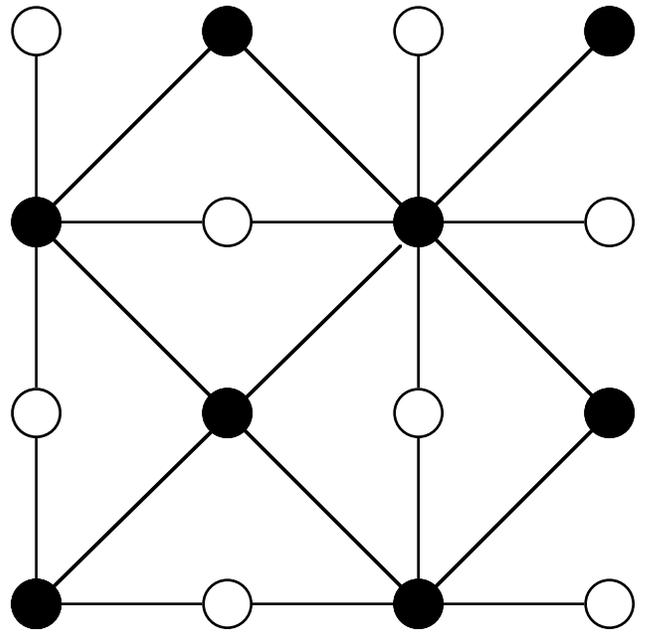
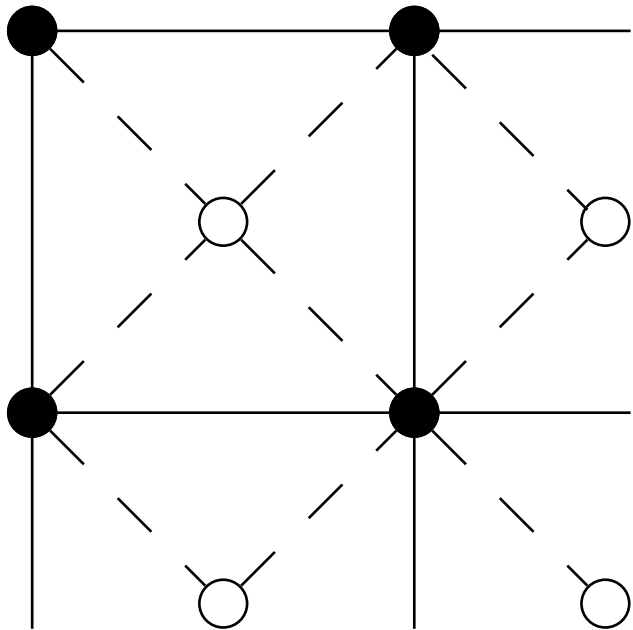
\end{document}